\documentstyle[12pt,titlepage]{article}
\def\baselinestretch{1.0}
\def\hf{\frac{1}{2}}

\def\bq{\begin{equation}}
\def\eq{\end{equation}}
\def\brr{\begin{eqnarray}}
\def\err{\end{eqnarray}}
\def\ba{\left(\begin{array}}
\def\ea{\end{array}\right)}

\def\ba{\left(\begin{array}}
\def\ea{\end{array}\right)}
\input epsf.tex
\epsfverbosetrue
\begin{document}
\pagestyle{empty}
\begin{flushright}
CERN-TH.7298/94
\end{flushright}
\ \vspace{1cm}
\begin{center} COSMOLOGY AND MODELS OF \\
SUPERSYMMETRY BREAKING\\ IN  STRING
THEORY \end{center}
\ \\
\begin{center}
Ram Brustein$^{*)}$
\\
Theory Division, CERN\\
 CH 1211 Geneva 23, Switzerland
\end{center}
\vspace{.5in}
\begin{center} ABSTRACT\end{center}
Supersymmetry breaking in string theory is expected to
occur when moduli  fields acquire non-trivial expectation
values.  In the early universe these
fields start out displaced from their final destinations. I present  
some recent
ideas about the cosmological evolution of the dilaton modulus field on  
the way
to its vacuum expectation value.\\
\vspace{1.4cm}\begin{center}
$^{*)}$Contribution to the proceedings of  SUSY94' Workshop,\\
May 14-171994, Ann Arbor, Michigan. \\ \end{center}
\noindent \rule[.1in]{16.5cm}{.002in}\\
\noindent \begin{flushleft} CERN-TH.7298/94 \\ June 1994
\end{flushleft}
\vfill\eject 
\setcounter{page}{1}
\pagestyle{plain}

Supersymmetry (SUSY) breaking in string theory is expected to
occur  when certain gravitationally interacting fields,
moduli, obtain vacuum expectation values (VEV's) \cite{dxl}.
These VEV's may break SUSY spontaneously. When considering the   
cosmological
time evolution of these fields an obvious question arises \cite{BS}.  
The
fields, because of thermal effects and quantum fluctuations, start out
displaced from their final destinations. How do
they get to their VEV's? Is there a reasonable dynamical
evolution that brought them there? It is not always possible
to ignore these questions, even by specifying   very
special initial conditions, because thermal fluctuations and
quantum fluctuations tend to smear the initial conditions.
 A partial list  of previous work on the subject is given in  
\cite{previous}.
The most interesting modulus field
seems to be the dilaton field,  $\phi$, whose expectation
value determines the string coupling parameter,
$g_{string}\sim <e^{\phi}>$.  The question seems to
arise for other moduli as well, for example, in the class of
models of the no-scale type \cite{fkz}.

A suggestion as to what the cosmological evolution
of the dilaton modulus field may be at the earlier stages was put  
forward
recently \cite{gv2,BV}. I combine these suggestions with ideas about  
the
evolution in the later stages as suggested in \cite{tv,BS,bkn,ccqr}.

$i$)  {\bf Accelerated Inflation}\\
The first, and best understood, stage of the evolution starts when the
dilaton is far away from its value today, deep in the weak-coupling  
region
($\phi\ll-1$). The Hubble parameter, $H$, is small. The evolution in  
this epoch
is determined by the vacuum solution of the string dilaton-gravity  
equations of
motion \cite{gv1}. After a period  whose length is determined by the
initial conditions, but has to be at least one second (a very long  
time!) to
solve the horizon and flatness problems, a strong curvature phase is  
reached.

$ii$) {\bf ``Branch Change"}\\
After the long period of accelerated inflation, the universe is much  
larger
than at the beginning.  Curvatures and kinetic  energies are of the  
order of
the string curvature and energy.
The correct dynamical description of this phase
should, therefore, be stringy in nature. If the value of the dilaton is  
small
throughout
this stage of evolution, dynamics can be described by classical string  
theory
in terms of a two-dimensional conformal field theory. This stage is not  
yet
well
understood. At the moment, the only  existing examples are not quite  
realistic
 \cite{kk,tsc}. In this epoch, a branch change, or a phase transition,
from the accelerated expansion phase into what  will eventually become  
a phase
of
decelerated expansion, occurs. This epoch lasts a short period and may  
be
identified with the  ``Big-Bang".
A large amount of radiation is generated
in this epoch. After some time the universe cools down enough and the  
next
stage
follows. The value which the dilaton takes at the end of this epoch is
important and determines many aspects of the later evolution.

$iii$) {\bf Radiation Domination}\\
Some time  after the ``branch change" event, the universe cools down  
and may
be
described accurately, again, by means of string dilaton-gravity   
effective
theory. Now, however, radiation and matter are important factors  
determining
the dynamical evolution. The dilaton remains approximately
at the  value that it had at the end of the ``branch change" epoch. The
universe evolves as a
regular Friedman-Robertson-Walker  (FRW) radiation-dominated universe.

$iv$) {\bf Dilaton Roll and Coherent Oscillations}\\
After a while the universe cools down further and  the dilaton  
potential
becomes important.  The dilaton starts to roll on its potential. To end  
up at a
non-trivial minimum the dilaton has to start this phase  in the  basin  
of
attraction of the minimum. As explained in \cite{BS}, for  
non-perturbative
potentials that are expected to induce SUSY breaking  the ``basin of
attraction" of their minimum is quite small.  Trapping the dilaton in a
minimum remains a  challenge for models of SUSY breaking.
If the dilaton gets into the basin of attraction of a non-trivial  
minimum it
coherently oscillates around the minimum, producing radiation in  
various forms,
and the universe reheats. Some aspects of this last stage were  
discussed in
\cite{bkn,ccqr,mg}.\\

I proceed to describe in more detail the different phases of  
cosmological
evolution. To describe the first phase, look for solutions of the   
effective
string  equations of motion in which the metric  is  of the isotropic,  
FRW type
with  vanishing spatial curvature\footnote{The situation for  
non-vanishing
spatial curvature is discussed in \cite{kne1}.}  and the dilaton  
depends only
on
time. \brr ds^2 &=& -dt^2+a^2(t) dx_i dx^i\nonumber \\ \phi&=&\phi(t)  
\err
The Hubble parameter, $H$, is related to  the scale factor,
$a$ in the usual way,
$H\equiv\frac{\hbox{\large $\dot a$}}{\hbox{\large $a$}}$.
Some algebra leads to three independent first order equations for the    
dilaton
and  $H$ \cite{gv1}. The original  dilaton equation  is a consequence  
of  these
equations, which read
\renewcommand{\theequation}{\arabic{equation}{a}} \bq
\dot H=\pm H\sqrt{3H^2+U + \hbox{\large $e^\phi$} \rho}-\hf U' + \hf
\hbox{\large $e^\phi$} p \setcounter{equation}{2}\eq
\renewcommand{\theequation}{\arabic{equation}{b}}\bq  
\setcounter{equation}{2}
\dot \phi= 3H\pm\sqrt{3H^2+U+\hbox{\large $e^\phi$} \rho}  \eq
\renewcommand{\theequation}{\arabic{equation}{c}}\bq\dot \rho + 3 H  
(\rho + p)
=
0 \setcounter{equation}{2}\label{fstordm}  \eq
\renewcommand{\theequation}{\arabic{equation}} Some sources in the form  
of an
ideal fluid were included \cite{gv1,vn} as well. The ($\pm$) signifies  
that
either a $(+)$  or $(-)$ is chosen for both  equations simultaneously.   
The
solutions to the equations (2a-2c) belong to two branches, according   
to which
sign is chosen. The $(+)$ branch has  some unusual  properties. In the  
absence
of any potential or sources the solution
for $\{H,\phi\}$  is  given by  \brr
 H^{(+)}&=&\pm \frac{1}{\sqrt{3}}\frac{1}{t-t_0}   \nonumber \\
\phi^{(+)}&=&\phi_0 + (\pm\sqrt{3}-1)\ln({t_0 -t}) ~~, ~~ t<t_0
\err
This solution describes either accelerated inflationary expansion
and  evolution from  a cold,   flat and weakly coupled universe towards  
a  hot,
curved and strongly coupled one  or accelerated contraction  and
evolution towards weak coupling. In general, the effects of a potential  
and
sources on this branch are quite mild. The dilaton of this branch   
flies
through
potential minima. As can be seen from Eq.(2b), if $H>0$,  it is  
impossible that
$\dot\phi=0$.   Inflation, in this solution, is driven,
 by the dilaton's kinetic energy, thanks to the negative value
($-1$) of the BD $\omega$ parameter\footnote{See \cite{lf} for a  
discussion of
these issues in a general Brans-Dicke theory.}. To solve the flatness  
and
horizon problems, at least 60 e-folds of inflation are  
necessary\footnote{There
may be subtleties associated with late time evolution which may change  
the
number 60 by some amount.}.
Setting $t_0=0$, one sees that
$t_{initial}/t_{final}{\
\lower-1.2pt\vbox{\hbox{\rlap{$>$}\lower5pt\vbox{\hbox{$\sim$}}}}\ }  
10^{43}$.
As we will see, this phase ends  when the curvature becomes strong.   
This
happens at $t_{final}/t_{string}\approx -1$. Therefore the minimal  
duration of
this stage is $10^{43} t_{string}\approx 1\ {\rm second}$. The dilaton
displacement during that time is  
$\sqrt{\alpha'}(\phi_{final}-\phi_{initial})>
(\sqrt{3}+1) \sqrt{3}\cdot 60 \sqrt{\alpha'}{\
\lower-1.2pt\vbox{\hbox{\rlap{$>$}\lower5pt\vbox{\hbox{$\sim$}}}}\ }  
300
\sqrt{\alpha'}$.

The last two stages of evolution are described by
the $(-)$ branch. This is a regular, negative feedback, branch.    The  
solution
$\{H^{(-)},\phi^{(-)}\}$  in the absence of potential or sources is  
simply
given
by
\brr  H^{(-)}&= \pm&\frac{1}{\sqrt{3}}\frac{1}{t-t_0}   \nonumber \\
\phi^{(-)}&=&\phi_0  +(\pm \sqrt{3}-1)\ln({t-t_0}) ~~, ~~ t>t_0 \err
This solution describes decelerated expansion ($H>0$, $\dot H<0$) or
decelerated contraction ($H>0$, $\dot H<0$) depending on the initial  
sign of
$H$. Correspondingly, the evolution is towards strong or weak coupling,
respectively. In the presence of appropriate sources, as in Eqs.(2),  
and when
the dilaton potential, $U$, is negligible the $(-)$ branch solution  
becomes an
ordinary FRW  expanding universe with constant dilaton (see also  
\cite{tv}).
The
radiation-dominated phase occurs when
\brr \hbox{\large $e^\phi$} \rho &\gg&
U(\phi) \nonumber\\ \hbox{\large $e^\phi$} \rho &\gg& U'(\phi)  
\nonumber\\
 p&=&\frac{1}{3}\rho \label{rd}
\err
Then, as a straightforward calculation shows, the solution of Eqs.(2)  
is
\brr
a&=&\sqrt{t}, \hspace{1in} H=\frac{1}{2t} \nonumber \\
\phi &=& \phi_{E} \nonumber \\
\rho &=& \frac{3}{2}
\hbox{\large $e$}^{\hbox{$-\phi_E$}} \frac{1}{a^4}
 \err
where $\phi_E$ is a constant.  During the branch change phase a
large amount  of radiation is expected to be produced. A rough estimate   
would
be $\rho\approx H_{\rm Branch Change}^4$, and therefore the conditions
(\ref{rd}) are expected to hold.

The branch change phase marks the transition from accelerated inflation  
to
decelerated expansion.  The transition between inflationary evolution  
and
a FRW expanding universe is usually referred to as
``graceful exit" from inflation and it is a well known problem to be  
faced  by
any model of inflation. The possibility  of graceful exit from  
accelerated
inflation is closely related, in our setup, to the question of whether  
the two
branches can be smoothly connected to one another. In \cite{BV} the  
possibility
of having a branch change while the curvature is still small was  
examined. The
conclusion was that this is not possible and that, therefore, if a  
branch
change
does occur, it has to be during a strong (string scale) curvature era.

It is not possible to continue and study the  evolution using the  
effective
dilaton-gravity theory because the whole field-theoretic framework  
breaks down.
If the dilaton VEV is still small  then classical string dynamics  
should be a
good approximation. The best is therefore to describe this phase using  
a 2-d
conformal field theory. An example of a conformal field theory that  
realizes a
branch change between two dual branches (similar, but not the same as  
the $(+)$
and $(-)$ branches) has been constructed \cite{kk,tsc},  and should be  
regarded
as an existence proof showing that such a transition can in fact occur  
during a
high curvature era. The branch change phase  is followed smoothly by a
radiation-dominated phase described by the $(-)$ branch as explained
previously.

\epsfysize=200pt
\centerline{\epsfbox{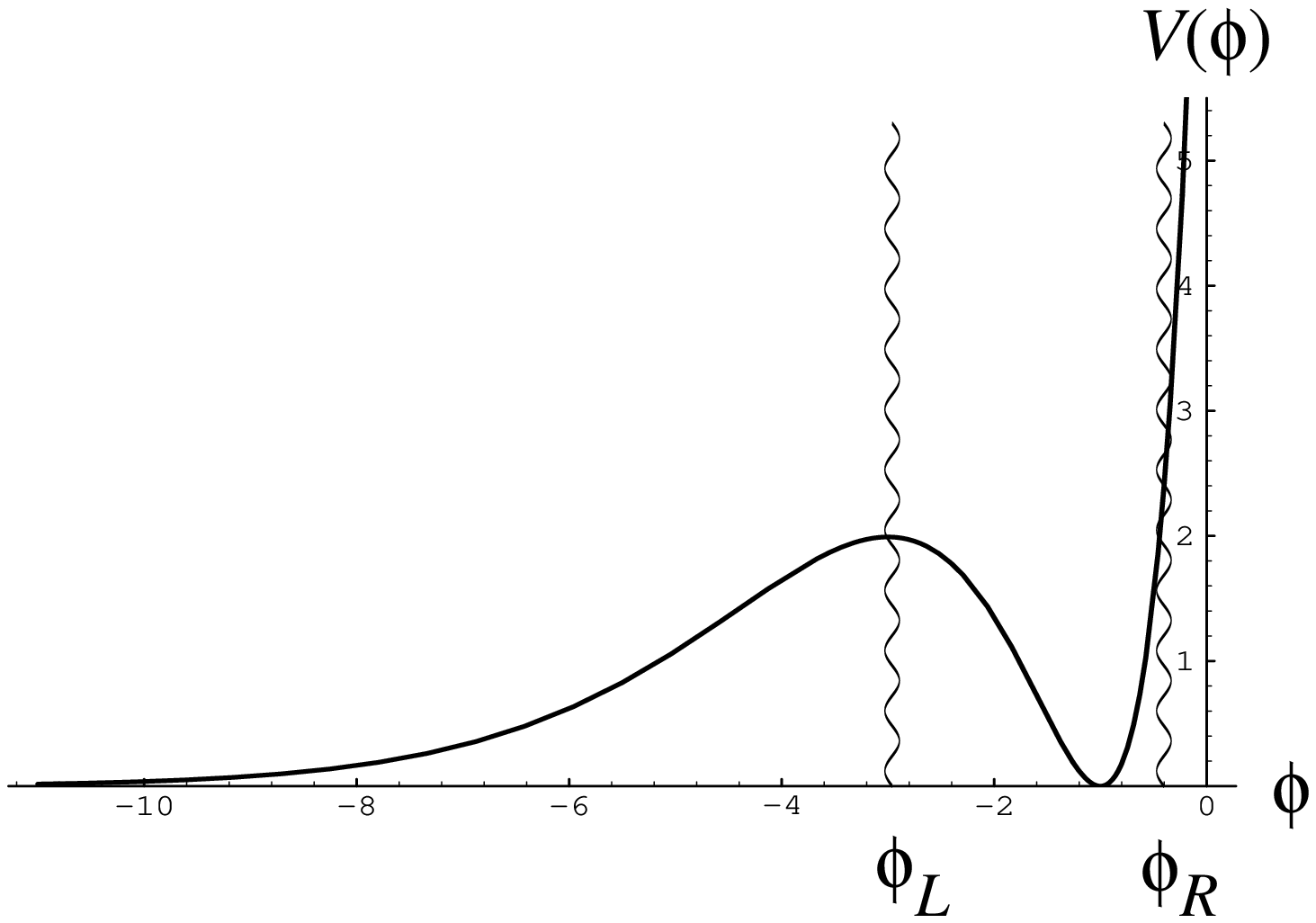}}

\renewcommand{\baselinestretch}{.5} {\small

\centerline{{\tenrm Figure 1. A typical expected dilaton potential.   
The wavy
lines  at $\phi=\phi_L$ and}} \centerline{{\tenrm \hspace{.60in}
$\phi\!=\!\phi_R$ mark the limits of the basin of attraction of the
 minimum  of }} \centerline{{ \tenrm
\hspace{.45in}     the potential.  The region of large negative values
of $\phi$ is the  weak }} \centerline{{ \tenrm \hspace{.10in} \    
coupling
region. The units on the vertical axis are
arbitrary. }} }

The  evolution during the fourth phase depends in a
stronger way on  the details  of the particular model of
supersymmetry breaking. It is in this phase that
cosmological evolution and models of supersymmetry breaking
are most closely related.  For a summary of relevant
properties of the dilaton and its interaction during this
phase see, for example  \cite{memor}. A typical expected dilaton
potential is shown in figure 1.

The exit value of the dilaton, $\phi_E$, determines its
subsequent evolution.  If $\phi_E<\phi_L$ or
$\phi_E>\phi_R$, the dilaton will continue to roll on its
potential  towards weak coupling \cite{BS}. After a
while the evolution will be  described by the vacuum $(-)$
branch. This is definitely not a description of  our
universe  today. If the exit value of the dilaton, $\phi_E$,
is within the basin of attraction of the minimum, marked by
the two wavy lines in figure 1, the dilaton will start to
coherently oscillate around the minimum and finally settle
down in its minimum. As it oscillates around the minimum
of the potential the dilaton  reheats the universe to a
temperature, $T_{RH}$, determined by the dilaton potential
and interactions. Different aspects of the coherent
oscillation phase have been considered in
\cite{bkn,ccqr,mg}.  It is clear that the deeper and wider
basin of attraction near the potential minimum, the better
the dilaton chances of getting trapped there.   The only
suggestion, to date, of how to explain the trapping of $\phi$ can be  
found in
\cite{hm}.

{\bf  Acknowledgement}\\
This talk is based in part on joint work with Paul Steinhardt and with
Gabriele Veneziano. I would like to thank Gabriele Veneziano for  
discussions.

\end{document}